\documentclass[a4paper,fleqn]{cas-dc}

\usepackage[numbers,sort&compress]{natbib}
\usepackage{amsmath,amssymb}
\usepackage{graphicx}
\usepackage{booktabs}
\usepackage{microtype}
\usepackage{multirow}

\ExplSyntaxOn
\RenewDocumentCommand \dashrule { O{.4pt} m m }
  {
    \par
    \color{black!50}
    \skip_vertical:n { #2 }
    \noindent \rule { \linewidth } { #1 } \par
    \normalcolor
    \skip_vertical:n { #3 }
  }
\RenewDocumentEnvironment { Abstract } { o }
  {
    \group_begin:
    \IfNoValueTF { #1 } { }
      { \tex_gdef:D \abstractname { #1 } }
    \parindent 0pt
    \box_if_empty:NTF \g_stm_key_box
      { \leftskip = .35 \textwidth }
      {
        \dim_gset:Nn \l_tmpa_dim { \box_ht:N \g_stm_key_box }
        \dim_gadd:Nn \l_tmpa_dim { \box_dp:N \g_stm_key_box }
        \leftskip .35 \textwidth
        \hspace* { -.35 \textwidth }
        \noindent \hbox_to_wd:nn { 0pt } { \box \g_stm_key_box \hss }
        \skip_vertical:n { - \l_tmpa_dim }
      }
    \noindent \abstractname \par
    \skip_vertical:n { -4pt }
    \noindent \rule { .65 \textwidth } { .2pt } \par \footnotesize
    \ignorespaces \everypar { \parindent=1.5em }
  }
  { \par \group_end: }
\ExplSyntaxOff

\makeatletter
\ExplSyntaxOn
\RenewDocumentCommand \printFirstPageNotes { }
  {
    \group_begin:
    \cs_if_exist:cT { H@@footnotetext }
      { \cs_set_eq:cc { @footnotetext } { H@@footnotetext } }
    \bool_if:NTF \g_stm_longmktitle_bool
      { \tex_let:D \columnwidth = \textwidth }
      { }
    \int_compare:nTF { \g_stm_jtype_int > 5 }
      { \stmaddress }
      { }
    \printtnotes
    \printnonumnotes
    \bool_if:NTF \g_stm_blind_bool
      { }
      {
        \printcornotes
        \printmaltese
        \printaddrinfoot
        \printemails
        \printurls
        \printfacebook
        \printtwitter
        \printgplus
        \printlinkedin
        \printfnotes
      }
    \bool_if:NTF \g_stm_longmktitle_bool
      {
        \if@twocolumn
          \tex_let:D \columnwidth = \Columnwidth
        \fi
      }
      { }
    \normalcolor
    \group_end:
  }
\ExplSyntaxOff
\makeatother

\makeatletter
\newcommand{\removecasfooter}{%
  \def\ps@cas{%
    \let\@oddhead\@empty
    \let\@evenhead\@empty
    \def\@oddfoot{\hfil\thepage\hfil}%
    \def\@evenfoot{\hfil\thepage\hfil}%
  }%
  \def\ps@plain{%
    \let\@oddhead\@empty
    \let\@evenhead\@empty
    \def\@oddfoot{\hfil\thepage\hfil}%
    \def\@evenfoot{\hfil\thepage\hfil}%
  }%
  \pagestyle{cas}%
}
\makeatother

\begin{document}
\removecasfooter
\thispagestyle{empty}

% ===================== Short title / short authors =====================
\shorttitle{IFPV}
\shortauthors{Zhigao Huang et~al.}

% ===================== Title =====================
\title[mode=title]{IFPV: An Integrated Multi-Agent Framework for Generative Operational Planning and High-Fidelity Plan Verification}

% ===================== Authors =====================
\author[1]{Zhigao Huang}
\ead{huangzg@gs.zzu.enu.cn}

\author[1]{Zhengqing Hu} 

\author[1,2,3,4]{Dong Chen}
\ead{chendongai@zzu.edu.cn}
\cormark[1]

\author[1]{Shaohan Zhang} 

\author[1,2,3,4]{Zhao Jin} 

\author[1,2,3,4]{Bo Zhang} 

\author[1,2,3,4]{Han Wu}

\author[1,2,3,4]{Mingliang Xu}
\ead{iexumingliang@zzu.edu.cn}
\cormark[1]

% ===================== Affiliations =====================

\affiliation[1]{organization={School of Computer and Artificial Intelligence, Zhengzhou University},
                city={Zhengzhou},
                postcode={450001},
                state={Henan},
                country={China}}

\affiliation[2]{organization={Engineering Research Center of Intelligent Swarm Systems, Ministry of Education},
                city={Zhengzhou},
                postcode={450001},
                state={Henan},
                country={China}}

\affiliation[3]{organization={National Supercomputing Center in Zhengzhou},
                city={Zhengzhou},
                postcode={450001},
                state={Henan},
                country={China}}

\affiliation[4]{organization={Henan Research Center for Large Model Technology and New Quality Software Engineering},
                city={Zhengzhou},
                postcode={450001},
                state={Henan},
                country={China}}           

\cortext[cor1]{Corresponding author.}

% ===================== Abstract =====================
\begin{abstract}
Operational plan generation and verification are critical for modern complex and rapidly changing battlefield environments, yet traditional generation and verification methods still respectively face the challenges of generation infeasibility and verification insufficiency. To alleviate these limitations, we propose an Integrated Multi-Agent Framework for Generative Operational Planning and High-Fidelity Plan Verification (IFPV).
IFPV consists of two tightly coupled modules: Multi-Perspective Hierarchical Agents (MPHA) for generative operational planning and an Adversarial Cognitive Simulation Engine (ACSE) for high-fidelity adversarial plan verification.
MPHA decomposes commander intent into executable multi-platform tactical action sequences through the collaboration of Pathfinder, Analyst, and Planner agents.
ACSE introduces an opponent equipped with a customized world model, which predicts the future evolution of mission-critical platforms and conducts dynamic counteractions against candidate plans.
Simulation experiments in the Asymmetric Combat Tactic Simulator (ACTS) show that IFPV improves mission success by 19.4\% and reduces operational cost by 41.7\% compared with a single-step large language model (LLM) planning baseline.
Compared with a traditional rule-based validator, ACSE increases the average suppression rate by 31.8\%, indicating that the proposed verification environment is stricter and more discriminative in revealing the latent vulnerabilities of candidate plans.
The code for IFPV can be found at \url{https://github.com/zhigao3ks/IFPV}
\end{abstract}

% ===================== Keywords =====================
\begin{keywords}
Generative planning \sep Large language models \sep Multi-agent system \sep World model \sep Adversarial verification \sep High-fidelity simulation
\end{keywords}

\maketitle
\thispagestyle{empty}

% ===================== Main text =====================

\section{Introduction}

Modern military operations are becoming increasingly complex, involving diverse combat units, heterogeneous weapon platforms, cross-domain operational effects, and rapidly changing battlefield situations.
In such environments, operational plan generation and verification are critical to mission success: the former transforms commander intent into executable tactical actions, while the latter determines whether a candidate plan can remain effective under dynamic confrontation.
Traditionally, operational plan generation has relied heavily on commander judgment, staff planning processes, and human operational experience, often following the Observe--Orient--Decide--Act (OODA) cycle \citep{boyd1996essence,price2023boyd}.
Meanwhile, plan verification has commonly depended on live exercises, manual wargaming, or rule-based simulation.
However, as intelligence data grows rapidly, physical and resource constraints become more tightly coupled, and decision windows become increasingly compressed, purely human-led plan generation and manually configured verification are no longer sufficient for high-intensity adversarial scenarios.
Therefore, automated operational plan generation and trustworthy plan verification have become increasingly important for modern battlefield decision support.

Reinforcement learning (RL) has shown considerable potential in sequential decision-making and adversarial strategy learning \citep{sutton2018reinforcement}.
Representative systems such as AlphaGo Zero, AlphaStar, and MuZero have demonstrated that self-play reinforcement learning, multi-agent reinforcement learning, and learned dynamics models can support strong decision-making in complex games and simulated environments \citep{silver2017alphagozero,vinyals2019alphastar,schrittwieser2020muzero}.
These studies indicate that data-driven agents can learn effective strategies through repeated interaction with simulated environments.
Nevertheless, RL-based methods typically rely on predefined state spaces, action spaces, and reward functions.
As a result, the learned policies are often tightly coupled with the training environment and may suffer from weak generalization when new platforms, weapon parameters, or deployment configurations are introduced.
In addition, RL training usually requires a large number of environment interactions, making it expensive to deploy in high-dimensional and rapidly changing combat scenarios.
For operational plan generation, these limitations can be summarized as \textbf{generation infeasibility}: existing automated generation methods often struggle to produce executable, resource-feasible, and coordinated multi-platform plans under complex battlefield constraints.

Beyond plan generation, reliable plan verification is equally critical, because a candidate plan that appears tactically reasonable at the textual or rule level may still fail under dynamic adversarial execution.
Existing plan verification methods mainly rely on constructive simulation systems, which derive scenario outcomes by presetting physical models, engagement rules, and behavioral scripts.
While such approaches offer good controllability and reproducibility, they still exhibit significant deficiencies in adversarial intelligence and verification fidelity \citep{banks2010discreteEvent,zeigler2000theory}.
On the one hand, opponent entities in these systems are often driven by static behavior trees, finite-state machines, or fixed scripts, and therefore lack the ability to dynamically counter candidate plans.
This may cause the verification process to degenerate into a one-sided target-shooting procedure.
On the other hand, traditional simulators are usually separated from plan-generation modules, and their outputs are often limited to coarse-grained indicators such as task completion, rather than providing discriminative feedback that can support plan selection and improvement.
For operational plan verification, these limitations can be summarized as \textbf{verification insufficiency}: existing verification methods often fail to provide sufficiently adversarial, high-fidelity, and informative verification for exposing latent vulnerabilities of candidate plans.

Large language models (LLMs) provide new opportunities for both automated operational planning and intelligent plan verification due to their strong prior knowledge, few-shot generalization, reasoning ability, and task decomposition capability \citep{brown2020language,wei2022chain}.
Studies have shown that LLMs can support reasoning-action interleaving, tree-structured deliberation, tool use, memory-based behavior simulation, and multi-agent collaboration in complex tasks \citep{yao2023react,yao2023tot,schick2023toolformer,park2023generativeAgents,wu2023autogen,wang2023voyager,guo2024multiAgentSurvey}.
These capabilities make LLMs promising not only for understanding unstructured mission descriptions and generating logically coherent tactical plans, but also for constructing role-based adversarial agents, interpreting simulation feedback, and supporting plan verification in dynamic environments.
However, real battlefield environments are not purely textual contexts.
They involve terrain, force deployment, platform states, weapon ranges, ammunition constraints, fire coverage, and temporal-spatial interactions.
General-purpose LLMs are not naturally equipped to precisely perceive these numerical and geometric constraints.
LLMs may also hallucinate unsupported content and still face substantial limitations in formal planning and reasoning about action effects \citep{ji2023hallucinationSurvey,valmeekam2022llmsCantPlan,kambhampati2024canLLMsPlan}.
These limitations affect both plan generation and plan verification: they may weaken the executability of generated plans and reduce the reliability of text-only evaluation.
For example, under long-context and multi-constraint conditions, LLMs may produce action conflicts, resource violations, physically infeasible maneuvers, or overconfident judgments that fail to reflect trajectory-level risks in high-fidelity simulation.

To alleviate these limitations, we propose an Integrated Multi-Agent Framework for Generative Operational Planning and High-Fidelity Plan Verification (IFPV).
IFPV is designed for complex dynamic battlefield environments and integrates commander intent parsing, candidate plan generation, adversarial simulation, and quantitative verification into a unified closed-loop workflow.
Specifically, IFPV consists of two tightly coupled modules.
The first module, Multi-Perspective Hierarchical Agents (MPHA), is responsible for generative operational planning.
It organizes Pathfinder, Analyst, Planner, and Validator agents into a structured workflow, where candidate routes are explored, tactical situations are assessed, multi-platform actions are coordinated, and hard constraints are checked before execution.
Through this collaborative hierarchy, MPHA transforms abstract commander intent into executable tactical action sequences while reducing path roughness, weak coordination, and infeasible actions commonly observed in single-step LLM planning.
The second module, the Adversarial Cognitive Simulation Engine (ACSE), is responsible for high-fidelity adversarial plan verification.
Built upon the Asymmetric Combat Tactic Simulator (ACTS), ACSE introduces an opponent equipped with a customized world model, which predicts the future trajectories of mission-critical platforms and conducts dynamic defense and firepower allocation accordingly.
To further improve this opponent's sensitivity to critical tactical situations, IFPV introduces Entity-Value-Aware Weighted Loss (EVA-Loss), which assigns larger training weights to high-value combat entities and guides the customized world model to prioritize mission-critical trajectories.
In this way, IFPV combines multi-agent plan generation, customized-world-model-based adversarial verification, and quantitative simulation feedback into a coherent framework for identifying executable, robust, and tactically meaningful plans.

The main contributions are summarized as follows:
\begin{itemize}
    \item We propose IFPV, an integrated multi-agent framework that unifies generative operational planning, high-fidelity adversarial simulation, and quantitative plan verification into a closed-loop workflow for complex dynamic battlefield environments.
    
    \item We design MPHA, a multi-perspective hierarchical agent architecture that decomposes commander intent into executable multi-platform tactical action sequences through the collaboration of Pathfinder, Analyst, Planner, and Validator agents.
    
    \item We construct ACSE, a high-fidelity plan verification module that introduces an opponent equipped with a customized world model into ACTS and incorporates EVA-Loss to improve the opponent's sensitivity to high-value combat entities and critical tactical situations.
    
    \item We validate IFPV through extensive simulation experiments. IFPV improves mission success by 19.4\% and reduces average operational cost by 41.7\% compared with a single-step large language model planning baseline, and increases the average suppression rate by 31.8\% compared with a traditional rule-based validator.
\end{itemize}

\section{Related Work}

\subsection{LLM-based Planning and Multi-Agent Collaboration}

Large language models (LLMs) have recently attracted increasing attention in complex task planning. Unlike traditional symbolic planning methods, LLMs usually do not require an explicitly constructed predicate-logic state space. Instead, they generate intermediate reasoning processes and action sequences from task descriptions through natural language understanding, contextual reasoning, and tool use. ReAct interleaves language reasoning traces with environment actions, enabling LLMs to update plans according to external feedback during execution \citep{yao2023react}. Tree of Thoughts (ToT) further extends single-path reasoning into multi-branch search, allowing models to self-evaluate, backtrack, and select among multiple candidate reasoning paths \citep{yao2023tot}. Recent surveys suggest that LLM-based planning methods can generally be categorized into task decomposition, plan search, external-module augmentation, reflection-based refinement, and memory-enhanced planning \citep{huang2024planningSurvey}.

Beyond these planning paradigms, chain-of-thought prompting has shown that explicitly generating intermediate reasoning steps can improve complex reasoning performance \citep{wei2022chain}, while Toolformer demonstrates that language models can learn to invoke external tools through API-style interactions \citep{schick2023toolformer}. Generative Agents further show that LLMs can be combined with memory, reflection, and planning modules to simulate believable agent behaviors \citep{park2023generativeAgents}. Despite these advances, recent studies on LLM planning indicate that LLMs still struggle with formal planning, action preconditions, and reasoning about state changes, especially when tasks require strict executability rather than textual plausibility \citep{valmeekam2022llmsCantPlan,kambhampati2024canLLMsPlan}.

Built upon these capabilities, LLM-based multi-agent systems (MAS) provide a more flexible organizational paradigm for solving complex tasks through role assignment, communication, negotiation, and tool augmentation. AutoGen introduces a multi-agent conversational framework for LLM applications, enabling multiple customizable agents to collaborate through natural language and code-based interaction \citep{wu2023autogen}. Voyager combines automatic curriculum learning, a skill library, and iterative prompting, allowing an embodied agent to continually acquire reusable skills in an open-world environment \citep{wang2023voyager}. Existing surveys also indicate that role design, communication protocols, memory mechanisms, and tool-use strategies are important factors affecting the quality of LLM-based multi-agent collaboration \citep{guo2024multiAgentSurvey}.

However, most existing studies on LLM-based planning and multi-agent systems focus on scenarios such as textual question answering, web navigation, code generation, open-world games, or general-purpose task collaboration. These scenarios usually involve weaker physical constraints, resource consumption, spatial geometry, and adversarial pressure than operational planning. For operational plan generation, a plan must be semantically reasonable while also satisfying low-level constraints such as platform maneuverability, weapon range, ammunition limits, suppression windows, and multi-platform temporal coordination. Directly using a single LLM to generate a complete plan may produce superficially coherent but physically infeasible action sequences. Similarly, loosely organized multi-agent dialogue may fail to align each role with the key decision stages of the operational planning workflow. To address these limitations, we propose MPHA, which explicitly assigns Pathfinder, Analyst, and Planner to path exploration, situation assessment, and global planning, respectively. Combined with rule-based verification and local retry mechanisms, MPHA improves both the executability and coordination quality of candidate plans.

Recent studies on small--large model collaboration and task-specific knowledge transfer further suggest that efficient planning systems need not rely solely on monolithic large models. Data Shunt proposes a collaborative paradigm in which small models and large models are dynamically routed according to the confidence of small models, reducing deployment cost while improving overall performance~\cite{chen2024data}. Its extended formulation further investigates how small models can improve large models and how large models can in turn benefit small models, forming a bidirectional collaboration framework for lower-cost and higher-performance inference~\cite{chen2025improving}. For decision-making tasks, Logic Distillation transfers decision logic from code function by function, showing that structured procedural knowledge can be distilled into models more effectively than simple output imitation~\cite{chen2025logic}. In addition, recent knowledge-injection studies extract task-relevant knowledge from large models to enhance downstream models, such as anomaly-related knowledge augmentation for vision anomaly detection~\cite{chen2025kka} and task-specific adaptation with lightweight small models under resource constraints~\cite{chen2025easy}. These studies are closely related to IFPV in motivation: rather than treating a general-purpose LLM as an all-in-one planner or validator, IFPV decomposes operational planning into specialized agent roles and injects entity-value-aware knowledge into the world model used by ACSE.

\subsection{World Models and Dynamic Environment Prediction}

World models aim to learn environment transition dynamics and support planning or decision-making by predicting future states. The idea of using learned models for planning can be traced back to integrated learning--planning--reacting architectures such as Dyna, where learned transition models are used to support simulated experience and policy improvement \citep{sutton1991dyna}. Later, world models were popularized as compact internal models that learn environment dynamics and support agent behavior \citep{ha2018worldmodels}. In model-based reinforcement learning (MBRL), world models are commonly used to simulate future trajectories, allowing agents to evaluate candidate actions in an imagined environment and thereby reduce the cost of real environment interaction. MuZero further demonstrates that planning with a learned model can achieve strong performance even without explicit knowledge of the environment dynamics \citep{schrittwieser2020muzero}. DreamerV3 demonstrates that learned environment models can support cross-domain generalization across diverse control tasks and improve policy learning through future-state prediction \citep{hafner2023dreamerv3}. In highly dynamic scenarios such as autonomous driving, generative world models such as GAIA-1 use video, text, and action inputs to model complex road environments, showing the potential of world models in future-scene generation, behavior prediction, and controllable simulation \citep{hu2023gaia1}. Recent studies have also explored whether LLMs can serve as implicit world models in complex environments, simulating the consequences of candidate actions and supporting model-based planning for agents \citep{gu2024webdreamer}.

Existing world-model studies mainly focus on robot control, autonomous driving, game agents, and web interaction, where the primary objective is often to improve the decision-making capability of the agent itself. In contrast, we employ a customized world model in the verification stage of operational plans, where it serves as the core predictive component of the opponent in ACSE. In other words, the world model is not used to directly help the plan generator produce better plans. Instead, it predicts the future trajectories of mission-critical platforms and supports dynamic defense and firepower allocation. Furthermore, we propose EVA-Loss, which introduces entity-level tactical value into the training objective of the world model. This design prevents the model from treating all entities uniformly and encourages it to prioritize high-value combat units that have greater impact on mission outcomes, thereby addressing the lack of entity-value awareness in general-purpose world models for operational scenarios.

\subsection{Intelligent Wargaming and Operational Simulation}

Intelligent wargaming and operational simulation are important research directions in military intelligence. Traditional wargaming and constructive simulation systems usually rely on manually defined rules, expert knowledge, and predefined behavior logic to conduct controllable simulation of combat processes. These systems offer good interpretability and reproducibility. However, when facing open-ended tactical variations and intelligent opponents, they are often limited by rule-base coverage and script complexity. In recent years, LLMs have been increasingly introduced into wargaming, military decision support, and adversarial simulation. Existing studies have discussed the methodological framework, application potential, and robustness of LLMs in wargaming \citep{chen2024llmWargaming}, while other works have constructed LLM- or vision-language-model-based multi-agent war simulation systems to emulate historical conflicts or complex battlefield interactions \citep{lin2024battleagent,hua2024waragent}.

These studies demonstrate the potential of LLMs in military simulation and complex adversarial modeling, but two limitations remain. First, many existing systems place more emphasis on macro-level narration, strategic reasoning, or role behavior simulation, while paying insufficient attention to low-level physical execution and weapon engagement mechanisms. As a result, they have limited ability to evaluate the execution quality of candidate plans along concrete spatiotemporal trajectories. Second, some studies directly use LLMs as simulation participants or validators, but they are not tightly integrated with high-fidelity simulation environments and quantitative metric systems. Consequently, their verification results may remain at the level of textual plausibility or subjective scoring.

Compared with existing work, the core distinction of IFPV lies in its simultaneous emphasis on high-quality generation and trustworthy verification. On the generation side, IFPV embeds the planning capability of LLMs into a structured operational plan generation workflow through MPHA, enabling candidate plans to explicitly satisfy spatial, resource, and multi-platform coordination constraints. On the verification side, IFPV advances plan verification from static rule checking to customized-world-model-driven adversarial verification through ACSE. Built upon ACTS, ACSE explicitly models entity movement, weapon range, hit randomness, suppression effects, and damage settlement, while using the customized world model to conduct real-time counteractions against candidate plans. In this way, IFPV forms a closed-loop framework of high-quality generation, high-pressure simulation, and quantitative feedback, providing a systematic solution for automated operational plan generation and trustworthy verification in complex dynamic battlefield environments.

\section{Methodology}

To address the difficulty of generating operational plans under complex constraints and the lack of intelligent adversarial verification in dynamic battlefield environments, we propose the Integrated Multi-Agent Framework for Generative Operational Planning and High-Fidelity Plan Verification (IFPV), an integrated generation--verification closed-loop framework. IFPV connects intent interpretation, candidate plan generation, adversarial simulation, and quantitative verification through unified data representation, executable interfaces, and feedback mechanisms. The system consists of two major components: Multi-Perspective Hierarchical Agents (MPHA) for plan generation and an Adversarial Cognitive Simulation Engine (ACSE) for high-fidelity verification. MPHA transforms abstract commander intent into executable tactical action sequences, while ACSE conducts dynamic stress testing of candidate plans in the Asymmetric Combat Tactic Simulator (ACTS). Accordingly, MPHA is designed to alleviate generation infeasibility, whereas ACSE is designed to alleviate verification insufficiency.

\begin{figure*}[pos=t]
\centering
\includegraphics[width=0.95\textwidth]{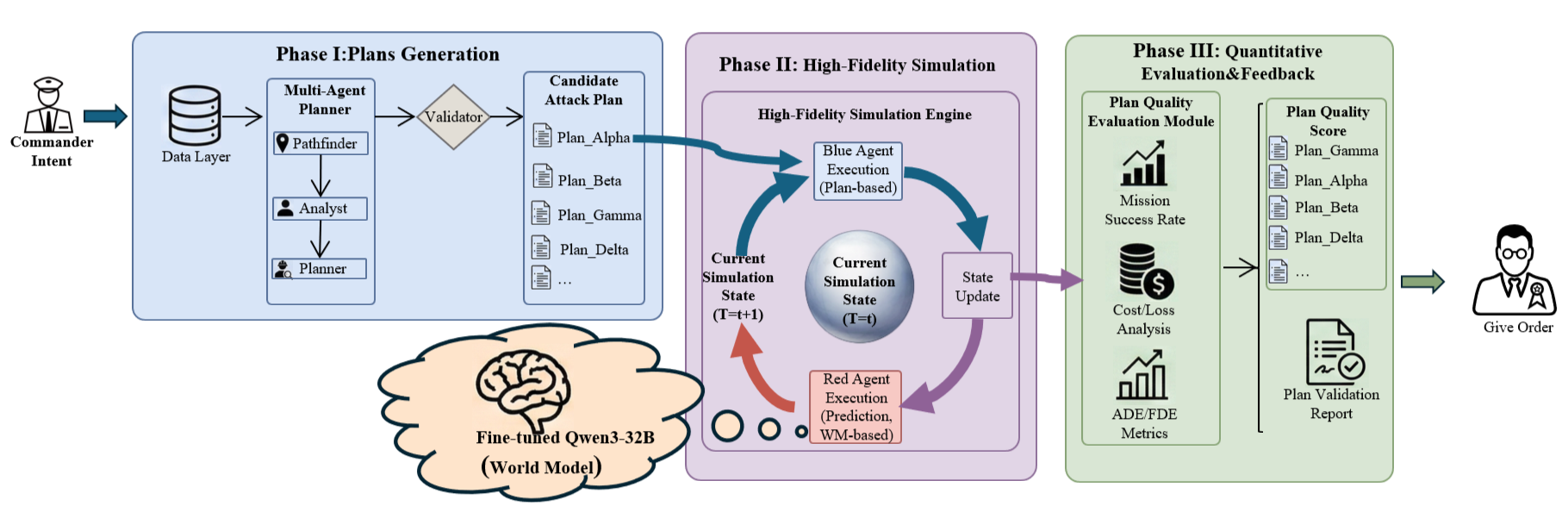}
\caption{End-to-end IFPV workflow for commander intent parsing, MPHA-based candidate plan generation, ACSE-based adversarial verification, and quantitative feedback.}
\label{fig:system_pipeline}
\end{figure*}

As illustrated in Fig.~\ref{fig:system_pipeline}, IFPV follows an end-to-end workflow. Given commander intent and structured battlefield information, MPHA first generates candidate operational plans. ACSE then imports these plans into ACTS, where the plan-executing side strictly executes the candidate plan and the opponent equipped with a customized world model performs dynamic counteractions. Finally, the evaluation module aggregates multiple simulation rollouts and produces a verification report based on mission success, operational cost, and trajectory deviation.

\subsection{Problem Formulation}

Let the natural-language mission issued by the commander be denoted as strategic intent $I$, and let the initial battlefield state be $S_0$. The state contains the positions, health values, ammunition, platform types, and weapon attributes of both the plan-executing side and the opponent side. The objective of IFPV is to generate a set of executable candidate plans under a global physical and resource constraint set $K$:
\begin{equation}
P = \{P_1, P_2, \ldots, P_n\}.
\label{eq:candidate_plan_set}
\end{equation}

Each candidate plan $P_i$ consists of timestamped atomic actions, such as waypoint maneuvering, weapon launch, suppression, and escort. The output format must be unambiguously executable by ACTS.

In the verification stage, a candidate plan $P^* \in P$ is imported into a discrete-time asymmetric game process. At time step $t$, the global system state is represented as
\begin{equation}
S_t = \langle X_t, H_t, A_t, E_t \rangle ,
\label{eq:global_state}
\end{equation}
where $X_t$ denotes the spatial coordinate matrix of entities, $H_t$ denotes health states, $A_t$ denotes remaining ammunition states, and $E_t$ denotes transient events such as firing, hits, and suppression.

Let $a_{\mathrm{blue}}^{(t)}$ denote the action set extracted from the candidate plan, $a_{\mathrm{red}}^{(t)}$ denote the opponent counteraction set generated by the customized world model, and $\xi^{(t)}$ be stochastic variables. The state transition of ACTS is formulated as
\begin{equation}
S_{t+1}
=
T_{\mathrm{sim}}
\left(
S_t,
a_{\mathrm{blue}}^{(t)},
a_{\mathrm{red}}^{(t)},
\xi^{(t)}
\right),
\label{eq:state_transition}
\end{equation}
where $T_{\mathrm{sim}}$ denotes the ACTS state transition function. Based on this formulation, the core task of IFPV is to generate high-quality candidate plans under $K$ and objectively evaluate their tactical effectiveness and robustness through ACSE.

\begin{table}[t]
\centering
\caption{Summary of key notations used in the methodology.}
\label{tab:notations}
\small
\begin{tabular}{@{}p{0.22\columnwidth}p{0.68\columnwidth}@{}}
\toprule
Symbol & Description \\
\midrule
$I$ & Commander intent \\
$S_t$ & Global system state at time step $t$ \\
$S_0$ & Initial battlefield state \\
$K$ & Global physical and resource constraint set \\
$P$ & Candidate plan set \\
$P^*$ & Candidate plan to be validated \\
$C$ & Set of macro-level courses of action \\
$T_{\mathrm{sim}}$ & State transition function of ACTS \\
$\omega_{i,j}$ & Entity-value-aware dynamic token weight \\
$PQS$ & Plan Quality Score \\
\bottomrule
\end{tabular}
\end{table}

\subsection{MPHA: Multi-Perspective Hierarchical Plan Generation}

\begin{figure*}[pos=t]
\centering
\includegraphics[width=0.95\textwidth]{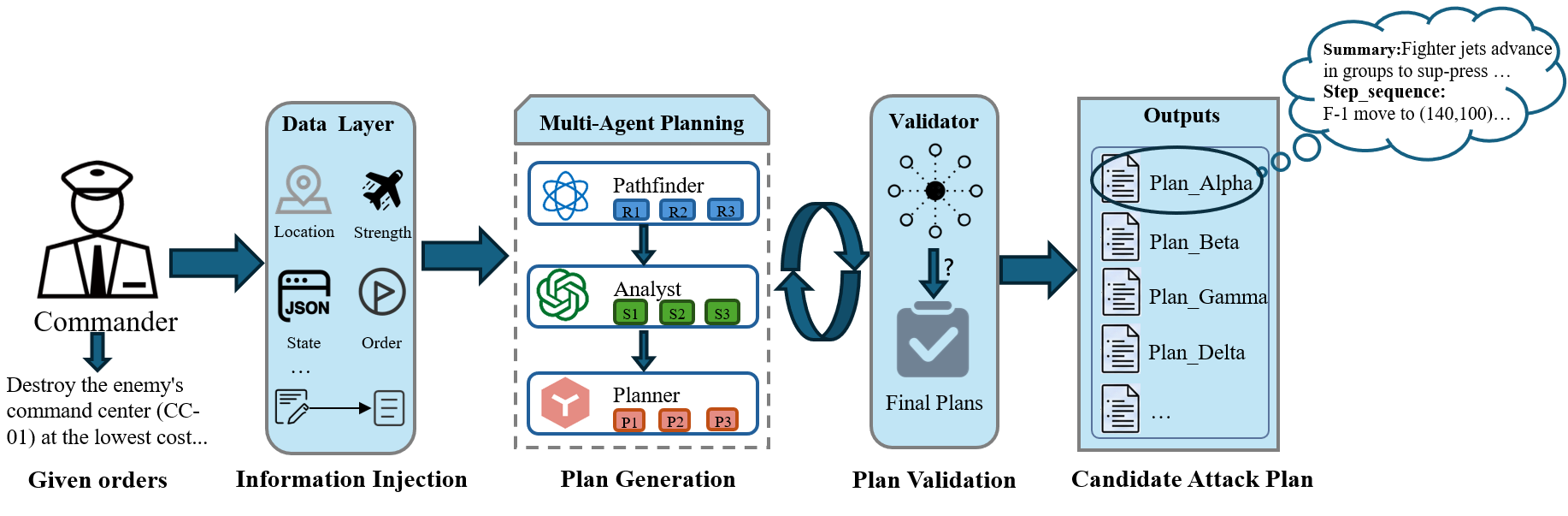}
\caption{Internal workflow of MPHA, including Pathfinder-based route exploration, Analyst-based situation assessment, Planner-based global coordination, and rule-based verification.}
\label{fig:mpha_workflow}
\end{figure*}

Operational plan generation is not a single-step text completion problem. It is a hierarchical decision-making process that involves geometric reachability analysis, situation-benefit assessment, and global resource coordination. Directly requiring a single LLM to generate a complete plan often leads to rough paths, action conflicts, resource violations, or weak coordination, which is consistent with prior observations on the limitations of LLMs in formal planning and reasoning about state changes \citep{valmeekam2022llmsCantPlan,kambhampati2024canLLMsPlan}. As shown in Fig.~\ref{fig:mpha_workflow}, MPHA decomposes plan generation into four stages: Pathfinder, Analyst, Planner, and Validator.

\subsubsection{Pathfinder}

Pathfinder generates a set of macro-level candidate courses of action from the perspective of global spatial geometry and threat distribution. Given commander intent $I$ and initial state $S_0$, it searches for route skeletons with geometric reachability and threat-avoidance potential:
\begin{equation}
C =
\operatorname{TopK}_{c \in \Omega_c}
R(c \mid I, S_0),
\label{eq:pathfinder}
\end{equation}
where $\Omega_c$ denotes the route search space, and $R(\cdot)$ measures the consistency between a candidate route and the commander intent, as well as its threat-avoidance benefit. Pathfinder does not directly generate the final plan, but provides geometrically safer and tactically promising route skeletons for subsequent agents.

\subsubsection{Analyst}

After macro-level candidate routes are obtained, Analyst performs fast forward simulation for each route and outputs a quantitative situation-assessment vector. For a candidate route $c_i$, the assessment result is formulated as
\begin{equation}
e_i =
F_{\mathrm{an}}(c_i,S_0)
=
[
\mathbb{E}(M_{\mathrm{success}}),
\mathbb{E}(L_{\mathrm{loss}}),
\mathbb{E}(T_{\mathrm{time}})
]^{\top},
\label{eq:analyst}
\end{equation}
where $M_{\mathrm{success}}$, $L_{\mathrm{loss}}$, and $T_{\mathrm{time}}$ denote estimated mission success, potential loss, and time cost, respectively. This assessment provides comparable intermediate evidence for Planner and prevents subsequent decisions from relying solely on coarse heuristics.

\subsubsection{Planner}

Planner receives the candidate route set $C$ and its corresponding assessment results $E=\{e_1,e_2,\ldots,e_k\}$, and generates executable multi-platform action sequences under the constraint set $K$. Its objective is
\begin{equation}
\begin{aligned}
P^*
=
\arg\max_{P \in \Omega_P(C,E)}
&\quad V_{\mathrm{global}}(P,E) \\
\mathrm{s.t.}
&\quad G(P) \leq K ,
\end{aligned}
\label{eq:planner}
\end{equation}
where $\Omega_P(C,E)$ denotes the plan space induced by candidate routes and assessment results, $V_{\mathrm{global}}(\cdot)$ denotes the global utility function, and $G(\cdot)$ denotes the constraint-checking function for resources, timing, and action logic. Planner is responsible not only for route selection, but also for coordinating task assignment, launch windows, and action timing among bombers, fighters, and suppression units.

\subsubsection{Validator}

Since LLMs may still produce local errors under strong numerical constraints, MPHA introduces a rule-based Validator after Planner. If a generated plan $P^{(r)}$ fails verification,
\begin{equation}
I_{\mathrm{valid}}\left(P^{(r)}\right)=0,
\label{eq:validator_check}
\end{equation}
the system extracts constraint-violation information $\Delta e^{(r)}$ and triggers local repair:
\begin{equation}
P^{(r+1)}
=
\Phi_{\mathrm{planner}}
\left(
P^{(r)},\Delta e^{(r)}
\right).
\label{eq:validator_repair}
\end{equation}
This planning--verification--repair loop improves the executability and consistency of candidate plans in the underlying simulator, thereby directly mitigating generation infeasibility in the plan-generation stage.

\subsection{ACSE: High-Fidelity Verification with a Customized-World-Model-Equipped Opponent}

After candidate plans are generated, IFPV enters the verification stage. Unlike traditional adversarial simulation based on static scripts or fixed rules, ACSE aims to construct a more dynamic and discriminative verification environment through an opponent equipped with a customized world model, following the broader idea that learned dynamics models can support prediction-centered planning and simulation \citep{sutton1991dyna,ha2018worldmodels,hafner2023dreamerv3}. ACSE does not simply increase opponent firepower parameters. Instead, it enables the opponent to generate targeted defensive strategies based on historical states and trajectory prediction, thereby exposing the potential vulnerabilities of candidate plans under realistic adversarial pressure. This design targets verification insufficiency by making the verification process more adversarial, dynamic, and informative.

\subsubsection{ACTS Simulation Mechanism}

ACSE runs on ACTS, a two-dimensional asymmetric air-ground combat simulator. ACTS handles entity state updates, weapon launch, damage settlement, and event logging, and explicitly models distance decay, hit randomness, and suppression effects. For a weapon with base hit probability $p_{\mathrm{base}}$, engagement distance $d$, and maximum effective range $R$, the effective hit probability is defined as
\begin{equation}
p_{\mathrm{eff}}
=
p_{\mathrm{base}}
\left(
\alpha
+
\beta
\left(
1-\frac{d}{R}
\right)
\right),
\quad d \leq R,
\label{eq:hit_probability}
\end{equation}
where $\alpha$ is the base hit-retention coefficient, $\beta$ is the distance-gain coefficient, and $\alpha+\beta=1$. When $d>R$, $p_{\mathrm{eff}}=0$. This mechanism reduces hit probability near the boundary of the engagement range and forces the planning system to balance penetration speed and strike accuracy.

ACTS also models firepower suppression. When an opponent-side air-defense unit is effectively attacked by support fighters, it enters a suppressed state for a duration $\tau_{\mathrm{sup}}$. Let $ROF_{\mathrm{base}}$ be its original firing interval. The effective firing interval and hit probability under suppression are updated as
\begin{equation}
ROF_{\mathrm{eff}}
=
ROF_{\mathrm{base}}
\times
\gamma_{\mathrm{rof}},
\label{eq:rof}
\end{equation}
\begin{equation}
p'_{\mathrm{eff}}
=
p_{\mathrm{eff}}
\times
\lambda_{\mathrm{hit}},
\label{eq:suppressed_hit}
\end{equation}
where $\gamma_{\mathrm{rof}}>1$ denotes the firing-rate penalty coefficient and $\lambda_{\mathrm{hit}}<1$ denotes the hit-probability decay coefficient. This mechanism allows fighters to provide soft-kill support and create short penetration windows for bombers.

\subsubsection{World Model Training with EVA-Loss}
\begin{figure*}[pos=t]
\centering
\includegraphics[width=0.95\textwidth]{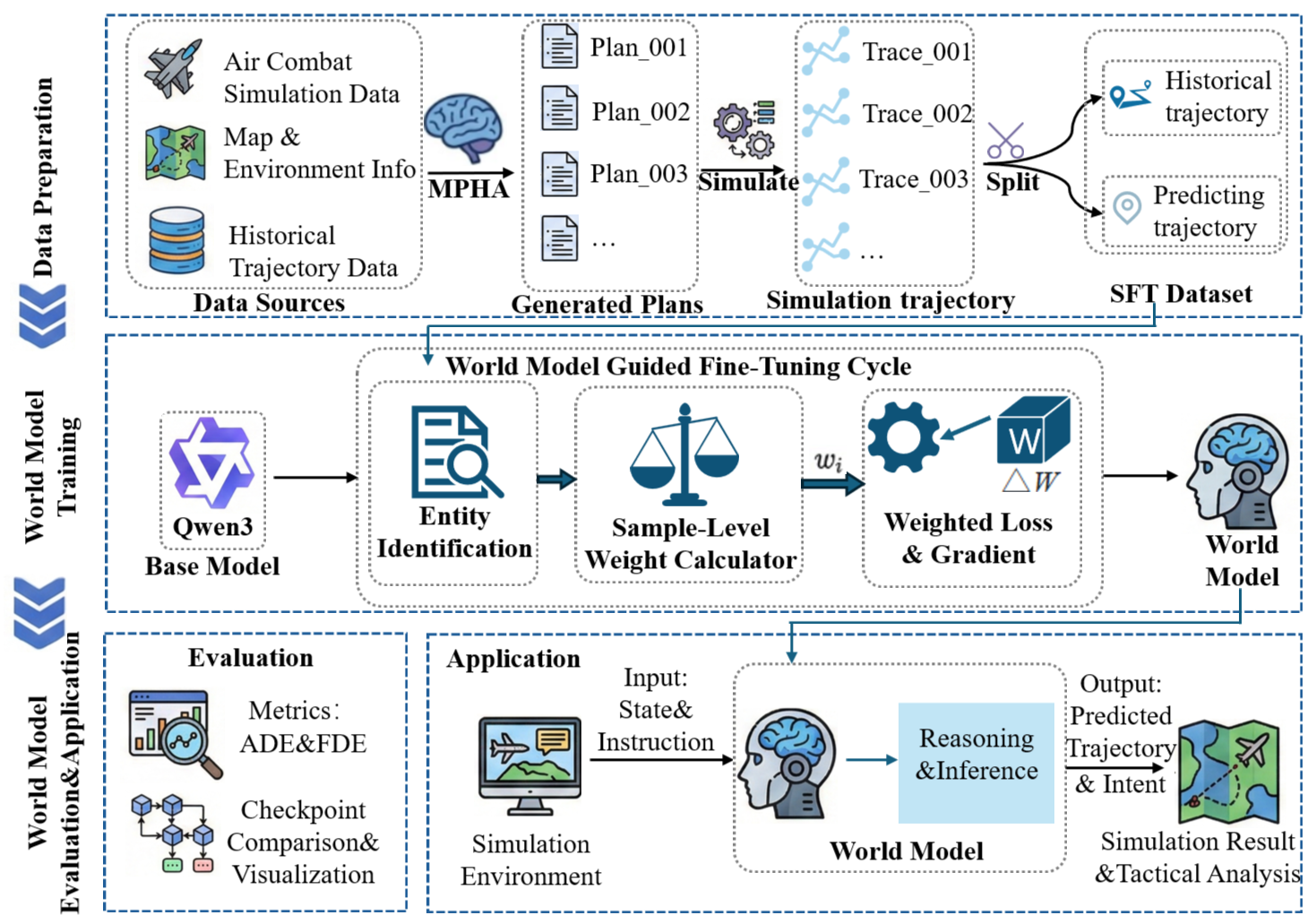}
\caption{World-model training, evaluation, and deployment pipeline for EVA-Loss-enhanced ACSE.}
\label{fig:world_model_pipeline}
\end{figure*}

To provide the opponent with state prediction and dynamic counteraction capability, we use Qwen3-8B as the base model and fine-tune it with battlefield trajectory data collected from ACTS. Qwen3 provides a family of dense and mixture-of-experts LLMs with strong reasoning and agent capabilities \citep{yang2025qwen3}. In our implementation, parameter-efficient adaptation is performed using LoRA, which freezes the base model parameters and injects trainable low-rank matrices into Transformer layers \citep{hu2021lora}. This design follows the broader idea that task-specific adaptation and knowledge injection can make large models more effective in specialized environments, while avoiding the cost of relying on a general-purpose model without domain grounding~\cite{chen2025improving,chen2025easy}. Unlike general knowledge-injection methods that mainly enhance downstream recognition or classification tasks~\cite{chen2025kka}, EVA-Loss injects tactical entity-value awareness into the world-model training objective, enabling the customized world model to prioritize mission-critical combat entities during trajectory prediction and counteraction generation. As shown in Fig.~\ref{fig:world_model_pipeline}, the world-model pipeline consists of simulation data collection, trajectory-prediction dataset construction, EVA-Loss-based fine-tuning, trajectory prediction evaluation, and ACSE deployment.

Let the historical observation sequence of the $i$-th sample be $X_i=S_{\leq t}^{(i)}$, and the future target sequence be $Y_i=S_{>t}^{(i)}$. The supervised fine-tuning dataset is constructed from ACTS rollouts.

Considering that high-value entities such as bombers have greater influence on mission outcomes, we propose Entity-Value-Aware Weighted Loss (EVA-Loss). Let $w_B$ and $w_F$ be the weights of high-value and ordinary entities, respectively, with $w_B>w_F$. For the $j$-th prediction token of the $i$-th sample, the dynamic weight is
\begin{equation}
\omega_{i,j}
=
w_B I_B(i,j)
+
w_F I_F(i,j),
\label{eq:eva_weight}
\end{equation}
where $I_B(i,j)$ and $I_F(i,j)$ indicate whether the current prediction position corresponds to a high-value entity or an ordinary entity. The final weighted supervised fine-tuning objective is
\begin{equation}
\mathcal{L}_{\mathrm{EVA}}
=
-\frac{1}{N}
\sum_{i=1}^{N}
\frac{1}{|Y_i|}
\sum_{j=1}^{|Y_i|}
\omega_{i,j}
\log
P
\left(
y_{i,j}
\mid
y_{i,<j},X_i;
W_0,\Theta_{\mathrm{LoRA}}
\right),
\label{eq:eva_loss}
\end{equation}
where $W_0$ denotes the frozen base-model parameters and $\Theta_{\mathrm{LoRA}}$ denotes the low-rank adaptation parameters. EVA-Loss guides the world model to focus more on trajectories of mission-critical combat units during training.

\subsubsection{Asymmetric Plan--Opponent Execution Logic}

During verification, the plan-executing side does not perform online re-planning, but strictly executes the candidate plan $P^*$ generated by MPHA. This ensures that the evaluated object is always the plan itself rather than additional online adaptation. At each time step $t$, the action set $a_{\mathrm{blue}}^{(t)}$ is extracted from the timestamped plan. Meanwhile, the opponent receives the historical state sequence $S_{\leq t}$ and generates counteractions using the fine-tuned world model:
\begin{equation}
a_{\mathrm{red}}^{(t)}
=
\arg\max_{a}
P_{\Theta_{\mathrm{WM}}}
\left(
a \mid S_{\leq t}
\right),
\label{eq:red_action}
\end{equation}
where $\Theta_{\mathrm{WM}}$ denotes the parameters of the fine-tuned world model. ACTS then updates the battlefield state according to Eq.~\eqref{eq:state_transition} and records trajectories, hit events, ammunition consumption, and platform losses.

\subsection{Quantitative Evaluation and Closed-Loop Feedback}

After multiple Monte Carlo simulation rollouts, the evaluation module assesses each candidate plan from three aspects: mission completion, resource cost, and execution robustness.

Mission Success Rate (MSR) measures whether the candidate plan successfully destroys the opponent's core target:
\begin{equation}
MSR(P^*)
=
\frac{1}{N}
\sum_{i=1}^{N}
I_{\mathrm{success}}^{(i)} ,
\label{eq:msr}
\end{equation}
where $I_{\mathrm{success}}^{(i)}$ indicates whether the $i$-th rollout succeeds.

Cost/Loss Analysis (CLA) jointly evaluates platform attrition and ammunition consumption:
\begin{equation}
CLA(P^*)
=
\frac{1}{N}
\sum_{i=1}^{N}
\left(
\eta_1 L_{\mathrm{friend}}^{(i)}
+
\eta_2 C_{\mathrm{ammo}}^{(i)}
\right),
\label{eq:cla}
\end{equation}
where $L_{\mathrm{friend}}^{(i)}$ and $C_{\mathrm{ammo}}^{(i)}$ denote platform attrition and ammunition consumption in the $i$-th rollout, and $\eta_1,\eta_2$ are cost weights.

Average Displacement Error (ADE) measures the deviation between simulated trajectories and planned trajectories:
\begin{equation}
ADE(P^*)
=
\frac{1}{NT}
\sum_{i=1}^{N}
\sum_{t=1}^{T}
\left\|
X_{\mathrm{sim}}^{(i)}(t)
-
X_{\mathrm{plan}}(t)
\right\|_2 .
\label{eq:ade}
\end{equation}

Finally, these metrics are integrated into a Plan Quality Score (PQS):
\begin{equation}
\begin{aligned}
PQS(P^*)
=
&\lambda_1 MSR(P^*)
-
\lambda_2 CLA(P^*) \\
&-
\lambda_3
\Phi_{\mathrm{norm}}
\left(
ADE(P^*)
\right),
\end{aligned}
\label{eq:pqs}
\end{equation}
where $\lambda_1,\lambda_2,\lambda_3$ are metric-fusion weights and $\Phi_{\mathrm{norm}}(\cdot)$ is a normalization function. The system ranks candidate plans according to $PQS$ and generates the final verification report. This report can assist commanders in selecting candidate plans and can also serve as feedback for further improving MPHA generation quality or ACSE verification strength.

\section{Experiments}

To systematically evaluate the proposed IFPV framework, the experiments are organized around three questions. First, we examine whether the ACSE-World Model can accurately capture battlefield spatiotemporal dynamics. Second, we evaluate whether MPHA can generate higher-quality candidate plans under a unified verification protocol, thereby testing whether generation infeasibility is reduced. Third, we investigate whether ACSE and EVA-Loss can improve the pressure strength and discrimination capability of plan verification, thereby testing whether verification insufficiency is reduced. Accordingly, this section presents the experimental setup, trajectory prediction results, plan-generation comparison, validator suppression evaluation, ablation studies, and a system-level case study.

\subsection{Experimental Setup}

All experiments are conducted in ACTS, a two-dimensional asymmetric air-ground combat simulation environment. ACTS divides the battlefield into a deployment region and an opponent defense region. The plan-executing side consists of bombers responsible for the main strike mission and fighters responsible for escort, containment, and suppression. The opponent deploys a layered air-defense network around the core target, including long-range interception units, medium- and short-range firepower units, and mobile air-defense entities. The purpose of ACTS is not to provide a static and low-pressure execution environment, but to expose the vulnerabilities of candidate plans under dynamic defense responses, stochastic engagement mechanisms, and asymmetric adversarial interactions.

Two difficulty levels are constructed on the same battlefield template: \textit{easy} and \textit{difficult}. The \textit{easy} setting corresponds to relatively sparse threat distribution and lower coordination pressure, while the \textit{difficult} setting involves denser fire coverage, more frequent dynamic patrols, and smaller tolerance margins for the plan-executing side. All experiments are repeated under the same initial-state template, entity parameters, and random protocol. During verification, the plan-executing side strictly executes the candidate plan without online re-planning, while the opponent reacts according to the assigned validator configuration.

For plan-generation evaluation, we report mission success rates in both \textit{easy} and \textit{difficult} scenarios. The overall success rate and robust success rate are defined as
\begin{equation}
S_{\mathrm{overall}}
=
\frac{S_{\mathrm{easy}}+S_{\mathrm{difficult}}}{2},
\label{eq:overall_success}
\end{equation}
and
\begin{equation}
S_{\mathrm{robust}}
=
\min(S_{\mathrm{easy}},S_{\mathrm{difficult}}).
\label{eq:robust_success}
\end{equation}
In addition, we report average platform attrition, average opponent fire hits, and average opponent fire fired to measure operational cost and engagement pressure. For validator strength, we use suppression rate as the main metric:
\begin{equation}
R_{\mathrm{sup}} = 1 - S ,
\label{eq:suppression_rate}
\end{equation}
where $S$ denotes the mission success rate under a given validator. A lower success rate and a higher suppression rate indicate a stricter verification environment.

\subsection{Trajectory Prediction of the World Model}

The ability of ACSE to construct a high-pressure verification environment depends on whether its world model can accurately predict the future trajectories of battlefield entities. Therefore, we first compare the fine-tuned ACSE-World Model with several general-purpose open-source models. All models are evaluated on the same test set, where the input consists of historical trajectories and current battlefield states, and the output is the future positions of combat entities. We use Average Displacement Error (ADE) and Final Displacement Error (FDE) as evaluation metrics, both of which are lower-is-better.

\begin{table}[t]
\centering
\caption{Trajectory prediction error comparison across general-purpose models and ACSE-World Model.}
\label{tab:trajectory_prediction}
\small
\resizebox{\columnwidth}{!}{
\begin{tabular}{lcc}
\toprule
Model & ADE $\downarrow$ & FDE $\downarrow$ \\
\midrule
GLM-4-9B & 18.01 & 22.67 \\
DeepSeek-R1-Distill-Llama-8B & 16.27 & 29.66 \\
DeepSeek-LLM-7B & 13.71 & 11.11 \\
Qwen3-32B (Base) & 6.05 & 8.46 \\
Meta-Llama-3-8B & 3.53 & 3.34 \\
Qwen2.5-7B & 2.72 & 3.55 \\
InternLM2.5-7B & 0.95 & 1.58 \\
ACSE-World Model & \textbf{0.18} & \textbf{0.54} \\
\bottomrule
\end{tabular}
}
\end{table}

As shown in Table~\ref{tab:trajectory_prediction}, general-purpose LLMs exhibit relatively large trajectory prediction errors in the battlefield setting. Even the best general baseline, InternLM2.5-7B, obtains an ADE of 0.95 and an FDE of 1.58. In contrast, ACSE-World Model reduces ADE and FDE to 0.18 and 0.54, respectively. Compared with InternLM2.5-7B, the reductions are approximately 81.1\% and 65.8\%. These results indicate that domain trajectory fine-tuning substantially improves the model's ability to capture local maneuvers, temporal transitions, and final tactical positions of combat entities.

\subsection{Comparison of Plan Generation Capability}

After confirming the trajectory modeling capability of the world model, we further compare the execution performance of different plan generators under the same ACSE verification environment. This experiment fixes the validator as ACSE so that the observed differences mainly reflect the quality of generated candidate plans.

\begin{table*}[t]
\centering
\caption{Comparison of plan quality among different plan generators under the unified ACSE verification environment. Higher overall and robust success rates, together with lower platform attrition and engagement pressure, indicate better plan quality.}
\label{tab:plan_generation}
\small
\resizebox{\textwidth}{!}{
\begin{tabular}{lccccccc}
\toprule
Plan Generator &
Overall Success $\uparrow$ &
Robust Success $\uparrow$ &
Easy Success $\uparrow$ &
Difficult Success $\uparrow$ &
Avg. Platform Attrition $\downarrow$ &
Avg. Opponent Fire Hits $\downarrow$ &
Avg. Opponent Fire Fired $\downarrow$ \\
\midrule
MPHA & \textbf{61.00} & \textbf{60.00} & \textbf{62.00} & \textbf{60.00} & 0.84 & 12.63 & \textbf{81.18} \\
Gemini 3.1 Pro & 49.33 & 48.00 & 48.00 & 50.67 & \textbf{0.75} & \textbf{12.61} & 82.04 \\
DeepSeek-V3 & 48.67 & 44.00 & 44.00 & 53.33 & 1.36 & 14.43 & 89.34 \\
GLM-5 & 13.33 & 0.00 & 26.67 & 0.00 & 2.25 & 17.56 & 110.72 \\
Qwen3-Max-Thinking & 2.67 & 1.33 & 4.00 & 1.33 & 2.29 & 20.14 & 127.80 \\
\bottomrule
\end{tabular}
}
\end{table*}

Table~\ref{tab:plan_generation} shows that MPHA achieves the best overall performance. It obtains 62.00\% success in the \textit{easy} setting and 60.00\% success in the \textit{difficult} setting, resulting in an overall success rate of 61.00\% and the highest robust success rate. Compared with the second-best generator, Gemini 3.1 Pro, MPHA improves overall success by 11.67 percentage points and robust success by 12.00 percentage points. These results indicate that the advantage of MPHA does not come from a single easy scenario, but from its ability to maintain balanced performance across different pressure levels.

In terms of operational cost, MPHA also maintains a favorable balance. Although Gemini 3.1 Pro has a slightly lower average platform attrition and opponent fire hits, the differences are marginal, while MPHA achieves clearly higher success rates and the lowest opponent fire fired. This suggests that MPHA does not rely on high-cost execution to complete the mission. Instead, its hierarchical structure enables more effective coordination between penetration, suppression, and strike actions. These results provide empirical evidence that MPHA alleviates generation infeasibility by improving executability, coordination, and cost control.

\subsection{Evaluation of Validator Suppression Capability}

The previous subsection demonstrates that MPHA can generate higher-quality plans under ACSE. However, this alone does not prove the effectiveness of the full generation--verification loop, because a weak validator may overestimate fragile candidate plans. Therefore, we further evaluate validator strength by comparing No\_Brain, GLM-5, and ACSE under the same candidate plan inputs. The goal here is not to determine which plan generator is better, but to examine which validator can more effectively reveal weaknesses in candidate plans.

\begin{table*}[t]
\centering
\caption{Verification results of different validators on representative plan generators. Lower success rate and higher suppression rate, platform attrition, opponent fire hits, and opponent fire fired indicate a stricter and more discriminative verification environment. For GLM-5-generated plans, only No\_Brain and ACSE are reported because using GLM-5 as both the plan generator and validator would introduce self-evaluation ambiguity.}
\label{tab:validator_suppression}
\scriptsize
\resizebox{\textwidth}{!}{
\begin{tabular}{lllccccc}
\toprule
Plan Generator & Difficulty & Validator &
Success Rate (\%) $\downarrow$ &
Suppression Rate (\%) $\uparrow$ &
Avg. Platform Attrition $\uparrow$ &
Avg. Opponent Fire Hits $\uparrow$ &
Avg. Opponent Fire Fired $\uparrow$ \\
\midrule
\multirow{6}{*}{Qwen3-Max-Thinking}
& \multirow{3}{*}{easy}
& No\_Brain & 6.67 & 93.33 & 2.05 & 12.20 & 64.71 \\
& & GLM-5 & 66.67 & 33.33 & 0.67 & 10.00 & 62.33 \\
& & ACSE & \textbf{4.00} & \textbf{96.00} & \textbf{2.11} & \textbf{16.64} & \textbf{103.56} \\
& \multirow{3}{*}{difficult}
& No\_Brain & 4.00 & 96.00 & 1.88 & 12.96 & 76.28 \\
& & GLM-5 & 33.33 & 66.67 & 2.00 & 21.33 & 138.33 \\
& & ACSE & \textbf{1.33} & \textbf{98.67} & \textbf{2.47} & \textbf{23.64} & \textbf{152.03} \\
\midrule
\multirow{6}{*}{Gemini 3.1 Pro}
& \multirow{3}{*}{easy}
& No\_Brain & 58.67 & 41.33 & 0.37 & 6.43 & 36.53 \\
& & GLM-5 & 100.00 & 0.00 & 0.25 & 6.92 & 40.17 \\
& & ACSE & \textbf{48.00} & \textbf{52.00} & \textbf{0.53} & \textbf{10.45} & \textbf{67.11} \\
& \multirow{3}{*}{difficult}
& No\_Brain & 54.67 & 45.33 & 0.13 & 6.01 & 31.65 \\
& & GLM-5 & 73.33 & 26.67 & 0.33 & 12.07 & 76.40 \\
& & ACSE & \textbf{50.67} & \textbf{49.33} & \textbf{0.96} & \textbf{14.76} & \textbf{96.96} \\
\midrule
\multirow{6}{*}{DeepSeek-V3}
& \multirow{3}{*}{easy}
& No\_Brain & 53.33 & 46.67 & 0.59 & 6.87 & 38.17 \\
& & GLM-5 & 53.33 & 46.67 & 0.80 & 9.80 & 67.27 \\
& & ACSE & \textbf{44.00} & \textbf{56.00} & \textbf{1.08} & \textbf{11.89} & \textbf{72.12} \\
& \multirow{3}{*}{difficult}
& No\_Brain & 65.33 & 34.67 & 0.55 & 8.27 & 54.97 \\
& & GLM-5 & 100.00 & 0.00 & 0.17 & 10.50 & 64.67 \\
& & ACSE & \textbf{53.33} & \textbf{46.67} & \textbf{1.63} & \textbf{16.97} & \textbf{106.56} \\
\midrule
\multirow{4}{*}{GLM-5}
& \multirow{2}{*}{easy}
& No\_Brain & 28.00 & 72.00 & 1.48 & 10.51 & 62.07 \\
& & ACSE & \textbf{26.67} & \textbf{73.33} & \textbf{1.97} & \textbf{14.80} & \textbf{89.69} \\
& \multirow{2}{*}{difficult}
& No\_Brain & 0.00 & 100.00 & 1.55 & 11.60 & 66.17 \\
& & ACSE & 0.00 & 100.00 & \textbf{2.52} & \textbf{20.31} & \textbf{131.75} \\
\bottomrule
\end{tabular}
}
\end{table*}

As shown in Table~\ref{tab:validator_suppression}, ACSE generally imposes stronger verification pressure across different plan generators and difficulty settings. For weak candidate plans generated by Qwen3-Max-Thinking, ACSE reduces success rates to 4.00\% and 1.33\% in the \textit{easy} and \textit{difficult} settings, respectively, while also increasing platform attrition and opponent fire pressure. Similar trends can be observed for Gemini 3.1 Pro and DeepSeek-V3, indicating that ACSE is not only effective for a specific generator type.

It is also worth noting that when candidate plans are already fully suppressed in terms of mission success, process-level metrics still provide additional information. For example, for GLM-5 in the \textit{difficult} setting, both No\_Brain and ACSE achieve a success rate of 0.00\%, but ACSE produces much higher opponent fire hits and opponent fire fired. This indicates that ACSE can further reveal the degree of exposure and firepower pressure during failed executions. Overall, ACSE improves verification strictness from three aspects: result-level suppression, attrition amplification, and process-level engagement pressure. These results show that ACSE alleviates verification insufficiency by providing more adversarial and more informative verification feedback.

\subsection{Ablation Study}

This subsection analyzes the sources of IFPV's performance gains from two perspectives. We first examine whether EVA-Loss improves the trajectory prediction and deployed verification capability of the world model. We then analyze the independent contributions of Pathfinder, Analyst, and Planner through MPHA module ablation.

\subsubsection{Ablation of EVA-Loss}

\begin{figure}[pos=t]
\centering
\includegraphics[width=0.95\columnwidth]{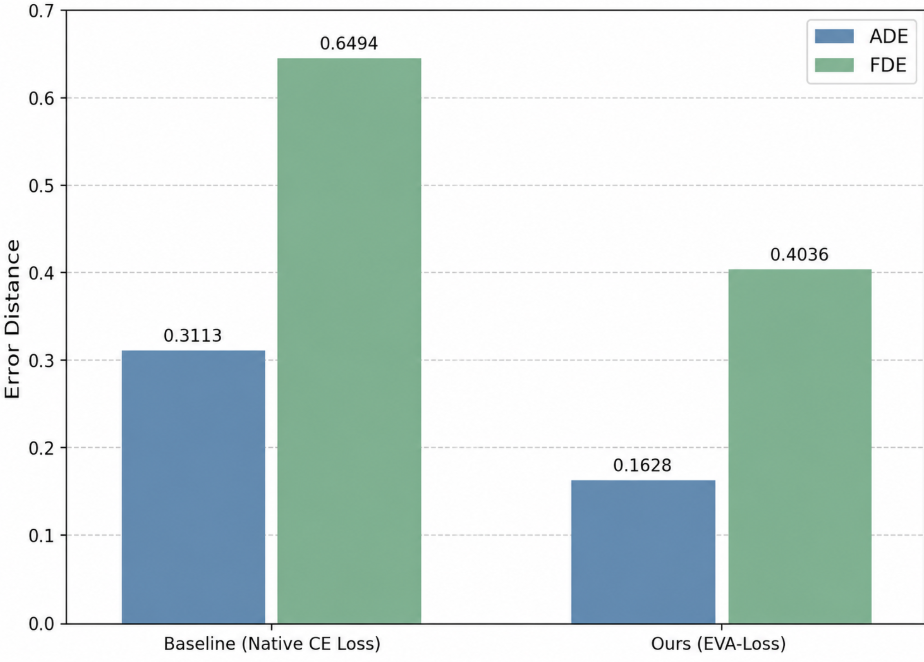}
\caption{Ablation comparison between EVA-Loss and Native CE Loss in world-model trajectory prediction.}
\label{fig:eva_loss_ablation}
\end{figure}

Fig.~\ref{fig:eva_loss_ablation} compares EVA-Loss with the native cross-entropy loss in trajectory prediction. With Native CE Loss, the world model obtains an ADE of 0.3113 and an FDE of 0.6494. After introducing EVA-Loss, ADE decreases to 0.1628 and FDE decreases to 0.4036, corresponding to reductions of approximately 47.7\% and 37.9\%, respectively. These results show that entity-value awareness enhances the model's ability to model trajectories of mission-critical combat units.

We further deploy world models trained with different losses into ACTS and evaluate seven representative candidate plans. As shown in Table~\ref{tab:eva_deployment}, ACSE (WM-EVA-Loss) generally achieves lower or comparable candidate-plan success rates across the representative plans, indicating that prediction accuracy improvements can be transferred into stronger dynamic defense. In particular, for Plan 1 and Plan 2, WM-NativeCE unexpectedly increases candidate-plan success rates to 90.0\% and 80.0\%, whereas ACSE (WM-EVA-Loss) suppresses both to 53.3\%. This suggests that the native world model may be misled by decoy entities, while EVA-Loss helps the customized world model focus on high-value threats such as bombers.

\begin{table}[t]
\centering
\caption{Candidate-plan mission success rates under three opponent configurations (\%).}
\label{tab:eva_deployment}
\small
\resizebox{\columnwidth}{!}{
\begin{tabular}{lccc}
\toprule
Candidate Plan & No WM & WM-NativeCE & ACSE (WM-EVA-Loss) \\
\midrule
Plan 1 & 62.0 & 90.0 & \textbf{53.3} \\
Plan 2 & 66.0 & 80.0 & \textbf{53.3} \\
Plan 3 & 84.0 & \textbf{50.0} & 53.3 \\
Plan 4 & 66.0 & 60.0 & 60.0 \\
Plan 5 & 68.0 & 70.0 & \textbf{66.7} \\
Plan 6 & 74.0 & 60.0 & 60.0 \\
Plan 7 & 78.0 & 80.0 & \textbf{73.3} \\
\bottomrule
\end{tabular}
}
\end{table}

\subsubsection{Ablation of MPHA Modules}

To analyze the contribution of internal MPHA modules, we compare the full MPHA with four ablated variants: \textit{single}, \textit{no\_pf}, \textit{no\_an}, and \textit{no\_pl}. The \textit{single} setting directly uses a single LLM to generate the final plan in one step. The \textit{no\_pf}, \textit{no\_an}, and \textit{no\_pl} variants remove Pathfinder, Analyst, and Planner, respectively. To reduce validator-induced strategy fluctuation, this experiment fixes the validator as No\_Brain.

\begin{table}[t]
\centering
\caption{MPHA module ablation results under a fixed No\_Brain validator.}
\label{tab:mpha_ablation}
\small
\resizebox{\columnwidth}{!}{
\begin{tabular}{lccccc}
\toprule
Metric & single & no\_pf & no\_an & no\_pl & MPHA \\
\midrule
Success Rate $\uparrow$ & 0.72 & 0.65 & 0.77 & 0.73 & \textbf{0.86} \\
TTK Mean (s) $\downarrow$ & 11.15 & 13.60 & \textbf{11.12} & 11.69 & 12.21 \\
Missiles Launched $\downarrow$ & 12.00 & 12.00 & 12.00 & 10.76 & \textbf{9.80} \\
Missile Hits $\uparrow$ & 4.54 & 4.32 & 4.58 & 4.61 & \textbf{4.76} \\
Missile Misses $\downarrow$ & 5.33 & 5.94 & 5.03 & 5.23 & \textbf{3.91} \\
Platform Attrition Mean $\downarrow$ & 2.28 & 2.41 & 1.43 & 1.97 & \textbf{1.33} \\
\bottomrule
\end{tabular}
}
\end{table}

Table~\ref{tab:mpha_ablation} shows that the full MPHA achieves the best overall performance in success rate, missile utilization, and platform-attrition control. Compared with \textit{single}, MPHA improves the success rate from 0.72 to 0.86, corresponding to a relative increase of approximately 19.4\%. It also reduces Platform Attrition Mean from 2.28 to 1.33, corresponding to a relative decrease of approximately 41.7\%. The \textit{no\_pf} variant obtains the lowest success rate and the highest number of missile misses, indicating that global path exploration is essential for high-quality planning. The \textit{no\_an} variant has the shortest TTK but a clearly lower success rate than MPHA, suggesting that removing situation assessment may lead to overly aggressive strategies. The \textit{no\_pl} variant exposes the importance of global temporal coordination and resource scheduling.

\begin{figure*}[pos=t]
\centering
\includegraphics[width=0.95\textwidth]{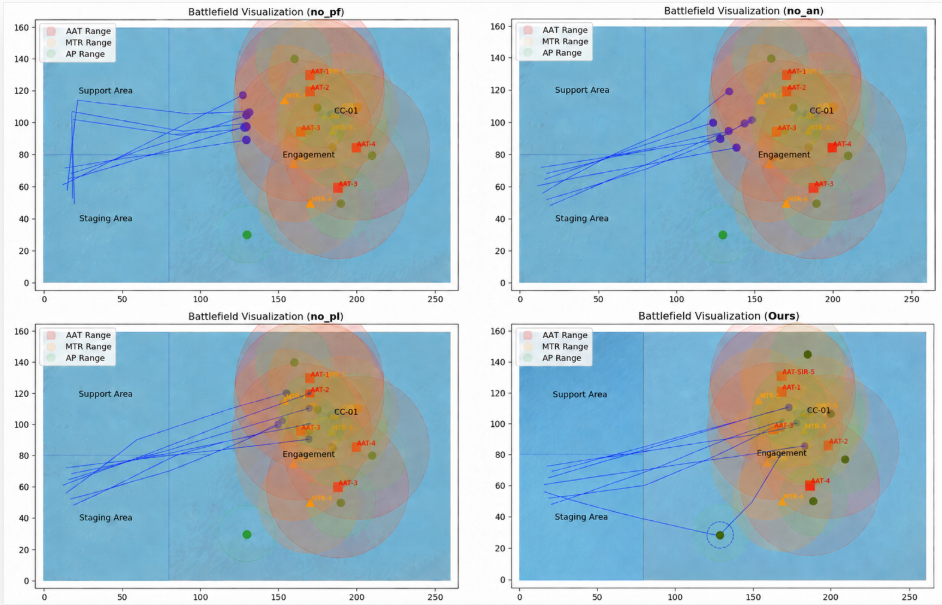}
\caption{Representative trajectories generated by ablated MPHA variants and the full MPHA. The \textit{no\_pf}, \textit{no\_an}, and \textit{no\_pl} variants are more likely to follow direct penetration patterns, whereas \textit{Ours} uses coordinated containment and covering maneuvers to support mission-critical strike units.}
\label{fig:mpha_ablation_trajs}
\end{figure*}

In addition to numerical metrics, we further visualize representative trajectories of the ablated variants and the full MPHA. As shown in Fig.~\ref{fig:mpha_ablation_trajs}, the ablated variants, including \textit{no\_pf}, \textit{no\_an}, and \textit{no\_pl}, tend to produce relatively direct penetration routes, premature convergence, or insufficient cross-platform coordination. In contrast, the full MPHA, marked as \textit{Ours}, forms a more tactical trajectory organization: one platform maneuvers to attract or contain patrol and air-defense attention, while more mission-critical strike units obtain safer penetration windows. These trajectory-level patterns explain why MPHA improves success rate, missile efficiency, and survivability.

It should be noted that MPHA does not achieve the shortest TTK. However, this does not imply lower efficiency. Instead, it reveals a more conservative and robust tactical style: MPHA sacrifices a small amount of execution time to achieve better firepower deconfliction, path avoidance, and cross-platform coordination. From an operational perspective, such a slightly slower but more reliable strategy is preferable to a faster but fragile plan.

\subsection{Case Study}

Finally, we select seven representative candidate plans generated by MPHA and compare LLM-based static ranking, human-expert ranking, and ACSE-based high-fidelity simulation ranking. The original static scoring considers tactical intent, multi-platform coordination, physical feasibility, and risk avoidance. For compact presentation, Table~\ref{tab:case_study} summarizes the static rankings and simulation rankings together.

\begin{table*}[t]
\centering
\caption{Comparison between static rankings and simulation-based rankings for seven representative candidate plans.}
\label{tab:case_study}
\small
\resizebox{\textwidth}{!}{
\begin{tabular}{lcccl}
\toprule
Plan & LLM Rank & Human Rank & Sim. Rank & Key Feedback \\
\midrule
Plan 1 & 3 & 1 & 2 & Strong concealment with limited opponent fire, but slightly lower penetration efficiency than Plan 6. \\
Plan 2 & 4 & 6 & 6 & Textually reasonable, but exposed to the opponent's predictive fire network and suffered severe losses. \\
Plan 3 & 5 & 5 & 5 & Mission completed, but a longer route increased exposure time and fire pressure. \\
Plan 4 & 2 & 3 & 3 & Standard tactical template with successful execution and reasonable engagement pressure. \\
Plan 5 & 6 & 4 & 4 & Underestimated by static scoring, but simulation shows good fire avoidance and escort effectiveness. \\
Plan 6 & 1 & 2 & 1 & Fast penetration and highest overall execution efficiency. \\
Plan 7 & 7 & 7 & 7 & Entered a cross-fire region and failed under concentrated opponent fire. \\
\bottomrule
\end{tabular}
}
\end{table*}

As shown in Table~\ref{tab:case_study}, static scoring and high-fidelity simulation may lead to different conclusions. For example, the LLM ranks Plan 2 fourth, but ACSE simulation shows that it fails under dynamic defense and suffers severe losses. Conversely, Plan 5 is ranked relatively low by the LLM, but simulation results show that it is executable and benefits from effective fire avoidance and escort behavior. Human-expert ranking is closer to simulation results, but it still cannot fully replace high-fidelity dynamic verification. For instance, experts rank Plan 1 above Plan 6, while ACSE identifies Plan 6 as the best plan in terms of overall execution efficiency.

The case study further confirms the necessity of high-fidelity dynamic verification. Static textual evaluation can provide preliminary screening based on tactical logic and expert intuition, but it cannot precisely quantify trajectory exposure, timing windows, and probabilistic fire coverage. By combining static reasoning with ACSE-based simulation, IFPV can identify pseudo-optimal plans that appear reasonable in text but are fragile under dynamic confrontation.

\section{Conclusion}

This paper proposed IFPV, an integrated multi-agent framework for generative operational planning and high-fidelity plan verification in complex dynamic battlefield environments. IFPV addresses the two central limitations identified in the introduction: MPHA alleviates generation infeasibility by decomposing commander intent into Pathfinder-based route exploration, Analyst-based situation assessment, Planner-based global coordination, and Validator-based constraint repair, while ACSE alleviates verification insufficiency by introducing a customized-world-model-equipped opponent into ACTS for dynamic adversarial stress testing. Experiments show that the fine-tuned ACSE-World Model substantially reduces trajectory prediction errors, MPHA achieves higher mission success and better operational cost control than single-step LLM planning baselines, and ACSE constructs a stricter and more discriminative verification environment than No\_Brain and GLM-5. The ablation studies and case analysis further confirm that the hierarchical design of MPHA and the entity-value awareness of EVA-Loss are both crucial for identifying executable, robust, and tactically meaningful plans under dynamic confrontation.

\section{Future Work}

Future work will proceed in three directions. First, the current ACTS environment mainly focuses on two-dimensional asymmetric air-ground confrontation, and it can be extended to three-dimensional space, multi-domain joint operations, and more complex terrain conditions to test the adaptability of IFPV under higher-dimensional physical constraints. Second, the current customized world model mainly relies on trajectory prediction and state-based counteraction; future work can incorporate intent recognition, deception detection, and multi-target priority modeling to further enhance the cognitive adversarial capability of ACSE. Third, interactive human-in-the-loop refinement can be explored so that IFPV forms a more natural human--machine collaborative loop among automatic generation, expert review, and simulation-based verification.

\section*{Declaration of competing interest}
The authors declare that they have no known competing financial interests or personal relationships that could have appeared to influence the work reported in this paper.

\section*{Data availability}
Data will be made available upon reasonable request.

% ===================== References =====================
\bibliographystyle{elsarticle-num}
\bibliography{references}

% ===================== Appendix =====================
\section*{Appendix}
%================================================
\appendix

\section{Experimental Parameters and Simulation Protocol}

This appendix provides detailed experimental parameters and simulation protocols used in the Asymmetric Combat Tactic Simulator (ACTS).

\subsection*{A.1 Scenario Configuration}

All experiments were conducted on a battlefield map of size $260 \times 160$.
Each scenario contains one command center (CC-01), multiple anti-air threat units (AATs), missile threat regions (MTRs), and air patrol (AP) units.

Two levels of tactical complexity were considered:

\begin{itemize}
    \item \textbf{Easy scenarios}: sparse threat deployment and limited overlapping fire coverage.
    \item \textbf{Difficult scenarios}: dense threat deployment with multiple overlapping interception regions.
\end{itemize}

Each candidate plan was evaluated under identical initial conditions.

\subsection*{A.2 Simulation Settings}

The main simulation parameters are listed below:

\begin{itemize}
    \item Simulation time step: $0.1s$
    \item Maximum simulation horizon: $20s$
    \item Number of Monte Carlo repetitions: $100$
    \item Random seeds: $\{1,\dots,100\}$
    \item Episode termination:
    \begin{itemize}
        \item Command center destroyed;
        \item All friendly units destroyed;
        \item Maximum simulation horizon reached.
    \end{itemize}
\end{itemize}

%================================================
\section{Static Scoring Rubric and Representative Plan Details}

This appendix presents the static scoring rubric used during candidate plan filtering, together with a representative end-to-end planning example generated by IFPV.

\subsection*{B.1 Static Evaluation Criteria}

Each candidate plan is evaluated according to:

\begin{itemize}
    \item Path smoothness;
    \item Threat avoidance;
    \item Resource consumption;
    \item Tactical coordination;
    \item Target engagement feasibility.
\end{itemize}

Each dimension is scored from 1 to 5, and the weighted sum is used for preliminary ranking before adversarial simulation.

\subsection*{B.2 Representative End-to-End Planning Case}

\begin{figure*}[htbp]
    \centering
    \includegraphics[width=0.95\textwidth]{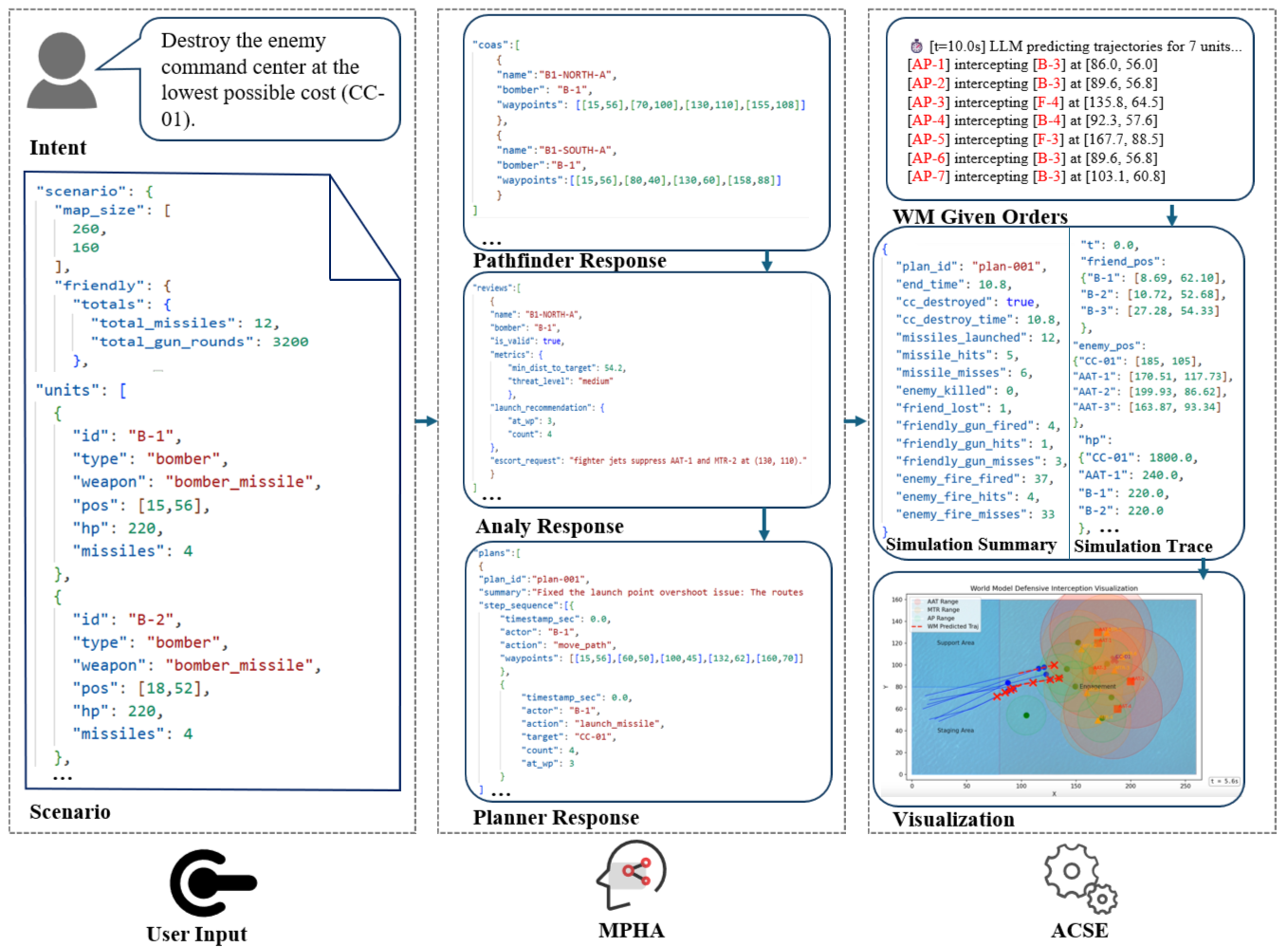}
    \caption{A representative end-to-end planning and verification case generated by IFPV.}
    \label{fig:appendix_pipeline}
\end{figure*}

Fig.~\ref{fig:appendix_pipeline} presents a representative operational-planning case, showing the complete workflow from commander intent parsing, scenario initialization, hierarchical multi-agent planning, world-model-based adversarial verification, and final battlefield visualization.

This example illustrates how MPHA alleviates the challenge of \textbf{generation infeasibility} by transforming high-level commander intent into executable multi-platform action sequences, while ACSE further alleviates \textbf{verification insufficiency} by introducing dynamic opponent behaviors and trajectory-level validation.

%================================================
\section{Training and Implementation Details}

This appendix provides implementation details of the customized world model.

\subsection*{C.1 Hardware and Software Environment}

All experiments were conducted using:

\begin{itemize}
    \item GPU: NVIDIA RTX A6000 (47.39 GB)
    \item CUDA: 12.8
    \item PyTorch: 2.10.0
    \item Transformers: 4.57.6
    \item PEFT: 0.18.1
\end{itemize}

\subsection*{C.2 Fine-Tuning Settings}

The customized world model was fine-tuned using LoRA.

\begin{itemize}
    \item Base model: Qwen3-8B
    \item LoRA rank: 16
    \item Learning rate: $2\times10^{-5}$
    \item Batch size: 8
    \item Number of epochs: 3
    \item Optimizer: AdamW
\end{itemize}

\subsection*{C.3 EVA-Loss Settings}

Entity-Value-Aware Weighted Loss (EVA-Loss) assigns larger training weights to mission-critical entities such as command centers and bomber platforms, enabling the customized world model to prioritize critical trajectory prediction and improving its sensitivity to tactical threats.

%================================================
\section{Metric Definitions and Calculation Details}

This appendix defines all evaluation metrics.

\subsection*{D.1 Plan Generation Metrics}

\begin{itemize}
    \item \textbf{MSR}: Mission Success Rate.
    \item \textbf{CLA}: Constraint-Level Accuracy.
    \item \textbf{PQS}: Plan Quality Score.
\end{itemize}

\subsection*{D.2 World Model Metrics}

\begin{itemize}
    \item \textbf{ADE}: Average Displacement Error.
    \item \textbf{FDE}: Final Displacement Error.
\end{itemize}

\subsection*{D.3 Simulation Metrics}

\begin{itemize}
    \item \textbf{Suppression Rate}:
\[
SR=
\frac{N_{hit}}{N_{fire}}
\]

    \item \textbf{Mission Success Rate}:
\[
MSR=
\frac{N_{success}}{N_{total}}
\]

    \item \textbf{Platform Attrition}:
\[
Attrition=
\frac{N_{lost}}{N_{friendly}}
\]
\end{itemize}

These metrics collectively demonstrate that IFPV effectively alleviates both \textbf{generation infeasibility} and \textbf{verification insufficiency} in complex dynamic battlefield environments.

\end{document}